\documentclass[journal,draftcls]{IEEEtran}
\ifCLASSINFOpdf
\usepackage{graphics,epsf,epsfig,fullpage,rotate,subfigure}
\else

\fi
\begin{document}
%
\title{\textsc{MaGe} - a {\sc Geant4}-based Monte Carlo framework for low-background experiments}
%
%
%

\author{Yuen-Dat Chan$^{1}$,
         Jason A. Detwiler$^{1,2}$,
	Reyco Henning$^{1,3}$, 
	Victor M. Gehman$^{2,4}$
	Rob A. Johnson$^{2}$, 
	David V. Jordan$^{5}$, 
	Kareem Kazkaz$^{2,6}$, 
	Markus Knapp$^{7}$, 
	Kevin Kr\"oninger$^{8,9}$, 
	Daniel Lenz$^{8}$,
	Jing Liu$^{8}$, 
	Xiang Liu$^{8}$, 
	Michael G. Marino$^{2}$, 
	Akbar Mokhtarani$^{1}$, 
	Luciano Pandola$^{10}$, 
	Alexis G. Schubert$^{2}$, 
	Claudia Tomei$^{10}$

\thanks{
  $^{1}$~Lawrence Berkeley National Laboratory, Berkeley, CA, USA, 
  $^{2}$~University of Washington, Seattle, WA, USA, 
  $^{3}$~University of North Carolina, Chapel Hill, NC, USA, 
  $^{4}$~Los Alamos National Laboratory, Los Alamos, NM, USA
  $^{5}$~Pacific Northwest National Laboratory, Richland, WA, USA	
  $^{6}$~Present address: Lawrence Livermore National Laboratory, Livermore, CA, USA, 
  $^{7}$~Physikalisches Institut, Universit\"at T\"ubingen, Germany, 
  $^{8}$~Max-Planck-Institut f\"ur Physik, Munich, Germany 
  $^{9}$~Present address: II. Institute of Physics, University of G\"ottingen, Germany,
  $^{10}$~INFN, Laboratori Nazionali del Gran Sasso, Assergi, Italy.
        }
	
} 

\markboth{Transactions on Nuclear Science (TNS)}%
{Detwiler \MakeLowercase{\textit{et al.}}: {\sc MaGe} - a GEANT4 based
Monte Carlo framework for low-background experiments}
%



\maketitle

\begin{abstract}
A Monte Carlo framework, \textsc{MaGe}, has been developed based on the
\textsc{Geant4} simulation toolkit. Its purpose is to simulate physics
processes in low-energy and low-background radiation detectors, specifically
for the \textsc{Majorana} and \textsc{Gerda} $^{76}$Ge neutrinoless double-beta decay 
experiments. This jointly-developed
tool is also used to verify the simulation of physics processes relevant
to other low-background experiments in \textsc{Geant4}.  
The \textsc{MaGe} framework contains simulations of prototype experiments and test stands, 
and is easily extended to incorporate new geometries and configurations 
while still using the same verified physics processes, tunings, and code framework.
This reduces duplication of efforts and improves the robustness of and confidence
in the simulation output.
\end{abstract}

\begin{IEEEkeywords}
Monte Carlo, neutrinoless double-beta decay, HPGe detectors, Geant4, radiation detection.
\end{IEEEkeywords}

%
\IEEEpeerreviewmaketitle


\section{Introduction}

{\sc MaGe} (MAjorana-GErda) is a {\sc
Geant4}-based~\cite{Agostinelli:2002hh,Allison:2006} Monte Carlo
framework jointly developed by the \textsc{Majorana}~\cite{Aalseth:2004yt} and
\textsc{Gerda}~\cite{Schoenert:2005} collaborations. Both experiments will
search for the neutrinoless double-beta decay ($0\nu\beta\beta$-decay)
of the $^{76}\mathrm{Ge}$ isotope using arrays of HPGe detectors. $0\nu\beta\beta$-decay is a
second-order weak process, the discovery of which is the only
practical way to determine if the neutrino is a Majorana particle (for
details, see the review article
in~\cite{Avignone:2007fu}). The purpose of
\textsc{MaGe} is to simulate the \textsc{Majorana} and \textsc{Gerda} 
experiments and their prototypes within a unified coding
framework. In the prototyping phase, the simulation
is used as a virtual test stand to guide detector design, to estimate the
effectiveness of proposed background reduction techniques, and to project
the experimental sensitivity. MaGe is used to develop detailed background models, 
and to study signal characteristics and systematic uncertainties. 
The simulation is also heavily employed in detector 
characterization and calibration tasks. The unified
framework allows the reuse of code and verified simulated physics
processes. The list of implemented physics models, the so-called
physics list in {\sc Geant4}, was optimized in \textsc{MaGe} for
low-background, underground physics
applications~\cite{Bauer:2004an}, with an emphasis on low-energy
interactions and hadronic interactions resulting from cosmic ray
spallation. \\



The  {\sc MaGe} concept is discussed in 
Section~\ref{section:concept}. The code structure
(Section~\ref{section:structure}) is followed by the implemented
physics list (Section~\ref{section:physics}). The code is validated by
the comparison of data from a variety of test stands and auxiliary
experiments with the results of accompanying simulations performed
with {\sc MaGe}. An example for the application of {\sc MaGe} is
presented in Section~\ref{section:example}. A summary of the
validation is given in Section~\ref{section:validation}.  Conclusions
are in the last section.


\section{Concept}
\label{section:concept}

The main goal of the development of {\sc MaGe} is the creation of a
robust, reliable and general-purpose Monte Carlo framework suitable for
the simulation of physics processes and background sources relevant
for low-background experiments, specifically $0\nu\beta\beta$-decay
experiments. \\

The structure of the framework allows (1) a parallel and independent
development of different branches of the code, for instance geometries
and interfaces; (2) ease of maintenance over the full life time of the
experiments; and (3) performing simulations with different
configurations by nonexperts. The flexibility of {\sc MaGe} stems from
the object-oriented nature of C++ and the {\sc Geant4} toolkit. The
{\sc MaGe} framework provides a single executable. The simulation can
be configured (geometry, I/O interface, etc.) with text file
macros. The macros are based on the {\sc Geant4} messenger which
allows users to run {\sc MaGe} in different configurations without
editing and recompiling the code. {\sc MaGe} can thus be run without
expert knowledge of the software without restricting its
flexibility. \\

The choice of {\sc Geant4} as the basis for {\sc MaGe} was motivated
by its flexibility and ambitious development within the particle and
medical physics communities. A wide number of physics models are
included and maintained in {\sc Geant4}. The most stringent
requirements for {\sc MaGe} are the proper simulation of the relevant
background sources for $0\nu\beta\beta$-decay
experiments. Specifically, this requires a precise description of
\begin{enumerate}
\item electromagnetic interactions from electrons and $\gamma$-rays at MeV and keV energies; 
\item radioactive isotope decay chains and nuclear de-excitation; 
\item interactions of thermal and fast neutrons; 
\item  the development of electromagnetic and hadronic showers
initiated by cosmic ray muons; 
\item penetration depths and ionization energy loss profiles of 
$\alpha$-particles. 
\end{enumerate}
{\sc Geant4} includes specific models for low-energy electromagnetic
physics~\cite{physics-manual}, for neutrons below 20~MeV, and for the
description of hadronic interactions resulting from cosmic ray
spallation. Furthermore, different models in the physics list,
tailored to fit specific physics applications (e.g. the simulation of
radioactive and muon-induced background contributions) can be selected
at runtime using macro commands. The same feature applies to many
important tuning parameters, such as the	 production cuts for $\delta$-rays and soft bremsstrahlung
photons. Specific details of the physics lists implemented in {\sc
MaGe} are discussed in Section~\ref{section:physics}. The
~\textsc{MaGe} code is regularly updated and ported in order to make
it compatible with the most recent {\sc Geant4} releases, which
include new functionalities, improvements and bug fixes. \\

A further feature of {\sc MaGe} is the existence  of interfaces to
other software. {\sc MaGe} interfaces to external event
generators and databases, the latter used to store information about detectors
and materials. Furthermore, waveform generators use the output of
{\sc MaGe} to calculate realistic electronic pulses from germanium detectors and hence provide
an end-to-end simulation of the experiments. Pulse-shape analysis (PSA) of these pulses allows
for powerful background reduction techniques. Monte Carlo simulations are used extensively
to test PSA algorithms and other such analysis routines, and to estimate their 
systematic uncertainties. An end-to-end simulation like {\sc MaGe} is also an 
indispensable tool for studying event topologies and problem cases.
{\sc MaGe} also has an abstract output interface that allows event data to be saved in ROOT~\cite{root}, AIDA-compliant~\cite{aida}, or text-based formats. Therefore {\sc MaGe}
maintains the ability to interface to external analysis tools.


\section{Structure of {\sc MaGe}}
\label{section:structure} 

The structure of {\sc MaGe} and {\sc Geant4} makes use of
the abstraction and object-oriented nature of C++. This allows the
user to select the required functionality at runtime by instantiating
a class that implements that functionality. The following
functionalities can be selected via {\sc Geant4} messengers at
runtime:
\begin{itemize}
\item \textbf{Physics Lists:} A collection of {\sc Geant4} physics
processes is called a physics list. They define the particles that are
included in the simulation and the decays and interactions they can
undergo.  There are several physics lists implemented in {\sc MaGe},
each optimized for the particular problem being simulated.  One list
is optimized at higher energies for simulating cosmic-ray muon
interactions, while others are optimized for standard electromagnetic
interactions at lower energies, i.e. below 10 MeV. These are used to
simulate the response of detectors to the decay of radioactive
isotopes. Each physics list is contained in its own class that is
instantiated at runtime. These lists may have unique associated
messengers to further refine the physics processes required.
\item \textbf{Geometries:} {\sc MaGe} currently has about 30
user-selectable geometries. Each geometry is encoded in a class that
derives from a base class that contains the basic components of a
geometry. The user can select a geometry at runtime via
messengers. This design also allows the reuse of existing geometry
classes, since the classes describing a geometry can be instantiated
within a class that requires that component. For example, a detailed
germanium crystal has been coded that is used many times in other
simulated detector geometries. This crystal can be simulated on its
own, or be instantiated many times in a complex detector array. The
use of the same basic geometry components eases coding and debugging
efforts.
\item \textbf{Output:} Each detector and Monte Carlo study has unique
output requirements. {\sc MaGe} has several different types of output
formats that can be combined to provide the information relevant to a
specific study. The base class for output is inherited by classes that
add functionality of a specific data analysis format, such as AIDA,
ROOT, or simple text-based output formats. These classes, in turn, are
inherited by classes that simulate and store detector responses and
save any relevant information. As for the geometries, basic components
of the output, such as the readout of a single crystal, can be
combined in a single class to create complex detector systems, such as
output for the entire \textsc{Majorana} or {\sc Gerda} detector arrays.
\item \textbf{Event Generators:} {\sc Geant4} provides a suite of
tools to create the initial conditions for an event. These include
radioactive decay-chain generators and simple volume samplers. {\sc
MaGe} also has the capability to use the {\sc Geant4} RDM~\cite{G4RDM} generator for radioactive
isotope decay. Furthermore it has a custom radioactive decay generator
for specific isotopes, such as \textsc{Decay0}~\cite{decay0}. This
allows for the later inclusion of such effects as angular correlation
between emitted $\gamma$-rays during $^{60}\mathrm{Co}$ decay, which is not
implemented in the RDM generator. \\ 

In addition, \textsc{MaGe} includes generators to simulate neutron and
muon backgrounds in underground laboratories, using either theoretical
models or data-driven approaches. Interfaces are available which read
initial conditions for an event from other codes, as
\textsc{Sources4A}~\cite{sources4a} for neutron flux and
\textsc{Musun}~\cite{musun} for muon flux. \\ 

Additional functionality was added in the form of a complex volume
sampler that can generate points uniformly distributed in any {\sc
Geant4} boolean solid. This is required to
simulate radioactive contamination embedded in detector components. A
surface sampler was also implemented that creates points uniformly
distributed on the surface of an arbitrary {\sc Geant4} solid~\cite{surfacesampler}. 
This is
required to simulate surface contaminations, in particular
$\alpha$-emitters. The user selects the appropriate generator at runtime
and the corresponding class is instantiated.
\item \textbf{Materials} {\sc MaGe} has the ability to read in all
relevant information about materials from a PostgreSQL database. This
is currently limited to quantities such as density, isotopic
abundance, etc. Once the \textsc{Majorana} or \textsc{Gerda} 
detectors are constructed,
the materials used will be carefully assayed and
characterized. All this information will be saved in a database as
well. {\sc MaGe} can then use this information to include the measured
activities in the simulation on a component-by-component level,
reducing systematic uncertainties in sensitivity calculations.
\end{itemize}

This design allows the simulation of detectors, prototypes and
validation experiments to be performed within the same framework using
the same physics processes, geometries and tools. This eases
cross-comparisons and reduces coding and debugging effort.


\section{Physics} \label{section:physics}

The physics list in {\sc MaGe} has been optimized for the reliable
simulation of the signal process and the most common background
sources in $0\nu\beta\beta$-decay experiments. It was selected
according to the suggestions of the \textsc{{\sc Geant4}}
team~\cite{physics_list} and optimized for
low-background physics
applications~\cite{Bauer:2004an,Pandola:2007}.\\

{\sc MaGe} has a default physics list that is mainly based on the
Underground Physics advanced example which is distributed with {\sc
Geant4}~\cite{physics_list}. The hadronic models implemented in the
physics list are

\begin{itemize}
\item theory-driven quark-gluon string models (QGSP) for pions, kaons
and nucleons with energies up to 100~TeV;
\item low energy parameterized (LEP) models~\cite{physics-manual}
for inelastic interactions of pions and nucleons with energies between
10 and 12~GeV, and for kaons below 25~GeV;
\item Bertini (BERT) or, alternatively, Binary (BIC) cascade models are used 
to describe nucleon and pion interactions below energies of 10~GeV;
\item data-driven neutron capture, fission, and elastic and inelastic 
scattering models from thermal energies to 20~MeV based on
tabulated cross-section data derived from the ENDF/B-VI
database~\cite{endf}  (HP models).
\end{itemize}
Alternative hadronic physics lists are available in \textsc{MaGe} that can 
be instantiated by messenger commands. Dedicated commands allow to use only LEP models 
(instead of cascades) for nucleons below 10~GeV, and to select 
Bertini or Binary models for nuclear cascades. For inelastic interactions of 
neutrons with energy below 8~GeV, it is also possible to use an alternative 
theory-driven quark-gluon string model (QGSC) which employs chiral-invariant 
phase-space modeling for nuclear de-excitation~\cite{physics-manual}. \\
Interactions of leptons with nucleons are simulated using the
equivalent photon approximation.  Photonuclear interactions
are modeled in detail and are divided into five energy regions:
\begin{itemize}
\item the giant dipole resonance region (10-30~MeV);
\item the quasi-deuteron region (from 30~MeV to the pion production threshold);
\item the $\Lambda$ region (from the pion production threshold to 450~MeV);
\item the Roper resonance region (450~MeV to 1.2~GeV);
\item the Reggeon-Pomeron region (1.2-3.5~GeV).
\end{itemize}
Hadronic final states are generated using a chiral-invariant phase-space decay model.
For energies above 3.5~GeV, photonuclear interactions are described by QGSP models.

Low-energy models~\cite{physics-manual} for the description of
electromagnetic interactions of $\gamma$-rays, electrons and ions
provided by {\sc Geant4} are used in {\sc MaGe} by default. These
models include atomic effects (e.g. fluorescence and Doppler
broadening) and can handle interactions down to energies of
250~eV. Synchrotron radiation is also included in the physics list for
electrons and positrons. Alternatively, electromagnetic interactions
of $\gamma$-rays, electrons and ions can be described by so-called
standard models provided by {\sc Geant4}. These models are tuned to
high-energy physics applications; they are less precise in the
low-energy region and do not include atomic effects. However, they are
faster in terms of computing time. The electromagnetic physics
processes provided by \textsc{Geant4} for $\gamma$-rays and $e^{\pm}$
(both standard and low-energy) have been systematically validated by
the \textsc{Geant4} Collaboration~\cite{Amako:2005} and by other
groups~\cite{Poon:2005} at the few-percent level. For the description
of electromagnetic interactions of muons, only standard models are
available in {\sc Geant4}. \\

{\sc MaGe} takes advantage of {\sc Geant4}'s ability to handle optical
photons. While the default {\sc MaGe} physics list does not include
interactions of optical photons these processes can be enabled during
runtime. The underlying models encompass scintillation light emission
(possibly with different light yields for electrons,
$\alpha$-particles and nuclei), Cherenkov light emission, absorption,
boundary processes, Rayleigh scattering and wavelength shifting. 
If optical photon treatment is enabled, it is
necessary to specify all relevant optical properties of interfaces and
bulk materials (refraction index, absorption length, etc.) in the
geometry definition. \\

{\sc Geant4} tracks all simulated particles down to zero range, although
various options exist to manually limit step size, track length, time-of-flight, 
and other parameters. Production cuts for
$\delta$-rays and for soft bremsstrahlung photons are
expressed in spatial ranges and are internally converted into energy
thresholds for the production of soft photons and $\delta$-rays in the
corresponding material. It is necessary to find a trade-off between
accuracy and computing time in most applications. Therefore {\sc MaGe}
provides three production cut realms: DarkMatter, DoubleBeta and
CosmicRays. The DarkMatter realm is used for high-precision
simulations, especially related to background studies for dark matter
applications: the cuts for $\gamma$-rays and betas 
are 5~$\mu$m and 0.5~$\mu$m, respectively, corresponding to
a $\sim$1~keV energy threshold in metallic germanium. The DoubleBeta realm
({\sc MaGe} default) is suitable for signal and background studies
related to double-beta decay, i.e. in the MeV energy-region: the
range cut for betas is lowered to 0.1~mm,
corresponding to a 100~keV threshold in metallic germanium. The
CosmicRay realm is used for the simulation of extensive
electromagnetic showers induced by cosmic ray muons. The
cut-per-region approach is used in this setup. Sensitive regions are
defined for which the production cuts are the same as for the
DoubleBeta realm. They are more relaxed everywhere else (5~cm for
$\gamma$-rays and 1~cm for betas). By avoiding the
precise tracking of particles in the inactive detector components,
computing time is saved. \\

     {\sc MaGe} includes some provisions to improve agreement between 
simulation and experimental results.  
For instance, simulations do not account 
for inefficient conversion of germanium nuclei recoil energy to 
ionization energy.  {\sc MaGe} contains output classes that simulate 
this conversion inefficiency. \\
 
{\sc Geant4} performs simulations on an event-by-event basis, where
each event begins with the release of a particle from a generator, and
ends when the interactions of the primary particle and its secondaries
have finished.  When long-lived radioactive decays occur, a single
{\sc Geant4} event may span many simulated years.  Output classes in
{\sc MaGe} divide {\sc Geant4} events into intervals that span user-selectable
times.  The total energy deposited during specific time intervals can
be reported.  This information can be used to simulate the
effectiveness of timing cuts at removing backgrounds.  It can also
improve agreement between results of {\sc MaGe} simulations and
experimental data.  Simulated energy deposits occurring long after the
duration of an experiment can be excluded.  Simulated energy deposits
in close succession can be summed to approximate the pile-up due to the finite time resolution
of data acquisition hardware used in an experiment. 


\section{Example: Liquid nitrogen test stand}
\label{section:example} 

The validation of the Monte Carlo code is an important step in the
development of {\sc MaGe}. This section presents the results of an
auxiliary measurement performed in the context of {\sc Gerda}. It is
an example of the validation procedure and shows the level of
agreement between data and Monte Carlo prediction for a particular
case. 

\begin{figure}[tbh]
\centering
\includegraphics[width=0.45\textwidth]{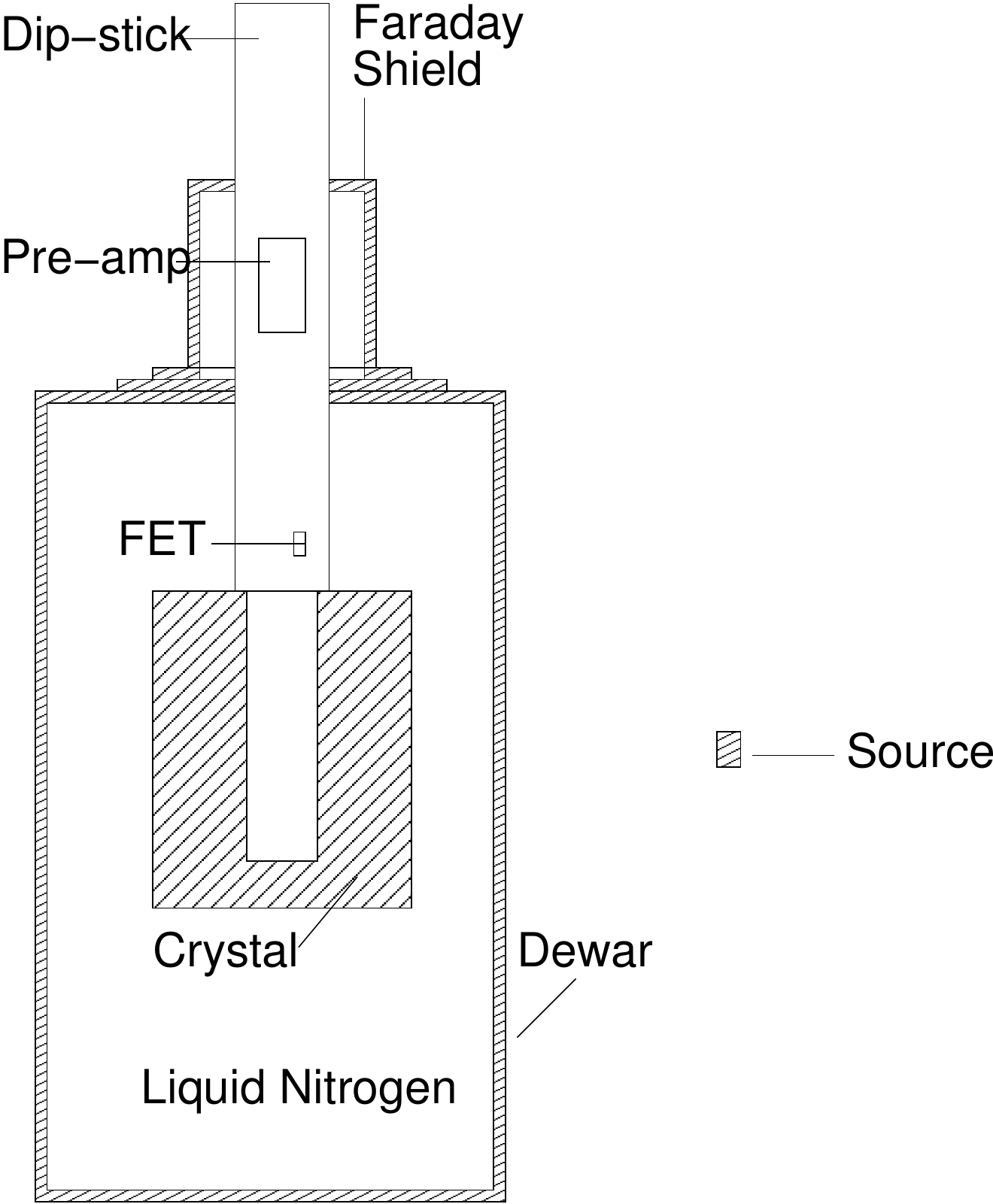}
\caption{Schematic drawing of the liquid nitrogen test stand. The
germanium detector is mounted onto a dip-stick and submerged into a
buffer of liquid nitrogen contained inside a double-walled aluminum
cryostat.
\label{fig:ln_teststand_scheme}} 
\end{figure}

The data used in the following comparison were obtained from a
measurement of radioactive $\gamma$-sources with a germanium detector
operated directly in a buffer of liquid nitrogen. The experimental
setup is shown in Fig.~\ref{fig:ln_teststand_scheme}. \\ 

The double-walled aluminum dewar containing the liquid nitrogen has a
diameter of 203~mm and a height of 410~mm. The two walls are separated
by 23.8~mm. The inner and outer wall thicknesses are 0.3~mm and
1.2~mm, respectively. The detector center is radially shifted by
17.7~mm from the center of the dewar. \\

The high-purity $n$-type germanium detector has a closed-ended coaxial
geometry. It has a height of 77.2~mm and a diameter of 64.5~mm. The
core electrode has a depth of 61~mm and a diameter of 10~mm. The dead
layer on the outer mantle is due to boron implantation and has a
thickness of about 0.3~$\mu$m. It is shielded by an 0.3~$\mu$m thick
aluminum layer. The inner dead layer is due to drifted lithium and has
a thickness of approximately 600~$\mu$m. The detector is operated at a
voltage of 3.5~kV. The FET of the pre-amplifier is operated close to
the detector inside the liquid nitrogen. The pre-amplifying
electronics are located outside the dewar and within a copper Faraday
shield. The DAQ energy threshold is set to 150~keV to avoid electronic
noise from the pre-amplifier. A software cut on the energy is applied
at 270~keV in order to avoid the low-energy region in which the
trigger efficiency is decreasing with decreasing energy. \\

Measurements were performed with three radioactive sources, $^{60}$Co,
$^{152}$Eu and $^{228}$Th. The sources were placed at a radial
distance of 10~cm from the dewar and vertically aligned with the center of the detector. An
additional background measurement was performed without a source
present. Around $1.4 \times 10^{6}$~events were collected in each measurement with a
source present (source data sets); around 150,000~events were
collected in the background data set. \\

The experimental setup was simulated using the {\sc MaGe}
framework built against {\sc Geant4} version 8.2 with patch01 applied. 
About $10^{8}$~events were simulated for each source. 
$^{60}$Co decays were generated using the {\sc Geant4} generator G4ParticleGun neglecting the angular correlation between the emitted $\gamma$-rays. 
Earlier studies have shown that in similar geometries no statistically significant differences between $^{60}$Co decays with and without angular correlation
between the emitted $\gamma$-rays were observed.
The energies obtained by the simulation were smeared according to the
energy resolution measured with the detector. The energy threshold
cuts applied to the measured data were also applied to Monte Carlo data. \\

\begin{figure}[tbh]
\centering
\includegraphics[width=1.0\columnwidth]{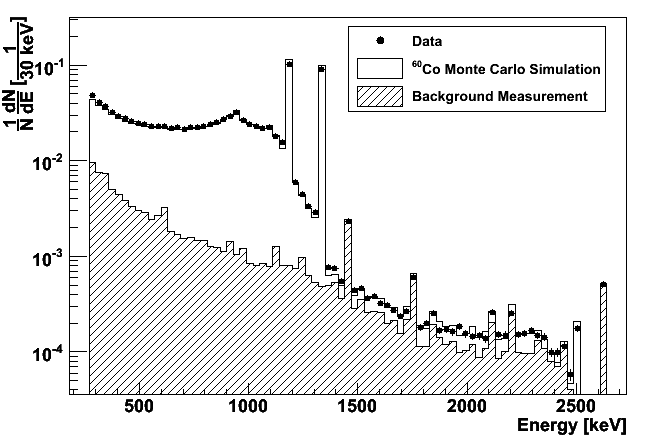}
\caption{Energy spectrum of the $^{60}$Co data set. The measured data
is represented by markers. The error bars show the statistical
uncertainty. The Monte Carlo and
background data sets are represented by the open and hatched
histograms, respectively.
\label{fig:datamcco60}} 
\end{figure}

In order to compare the measured data with Monte Carlo predictions the
background from radioactivity in the laboratory is estimated for each
source data set by scaling the background data set according to fits
the number of events in characteristics $\gamma$-lines.
The procedure is described in~\cite{Abt:2007rg}. 
%
%
As an example the energy spectrum obtained from the $^{60}$Co data set
is shown in Fig.~\ref{fig:datamcco60} together with the Monte Carlo-plus-background 
spectrum. The measured Compton continuum below 1.1~MeV
is well described by the simulation. The average deviation in that
region is approximately 5\% with the data being systematically higher
than the Monte Carlo plus background spectrum. The numbers of events
under the two characteristic peaks of $^{60}$Co at $1,173$~keV and
$1,332$~keV are higher in Monte Carlo compared to measured data by
about 10\%. The spectrum is dominated by background events for higher
energies. The disagreement between the measured and simulated data in this region is
less than 15\% on average with the data being systematically lower
than the Monte Carlo plus background spectrum. \\

\begin{figure}[tbh]
\centering
\includegraphics[width=1.0\columnwidth]{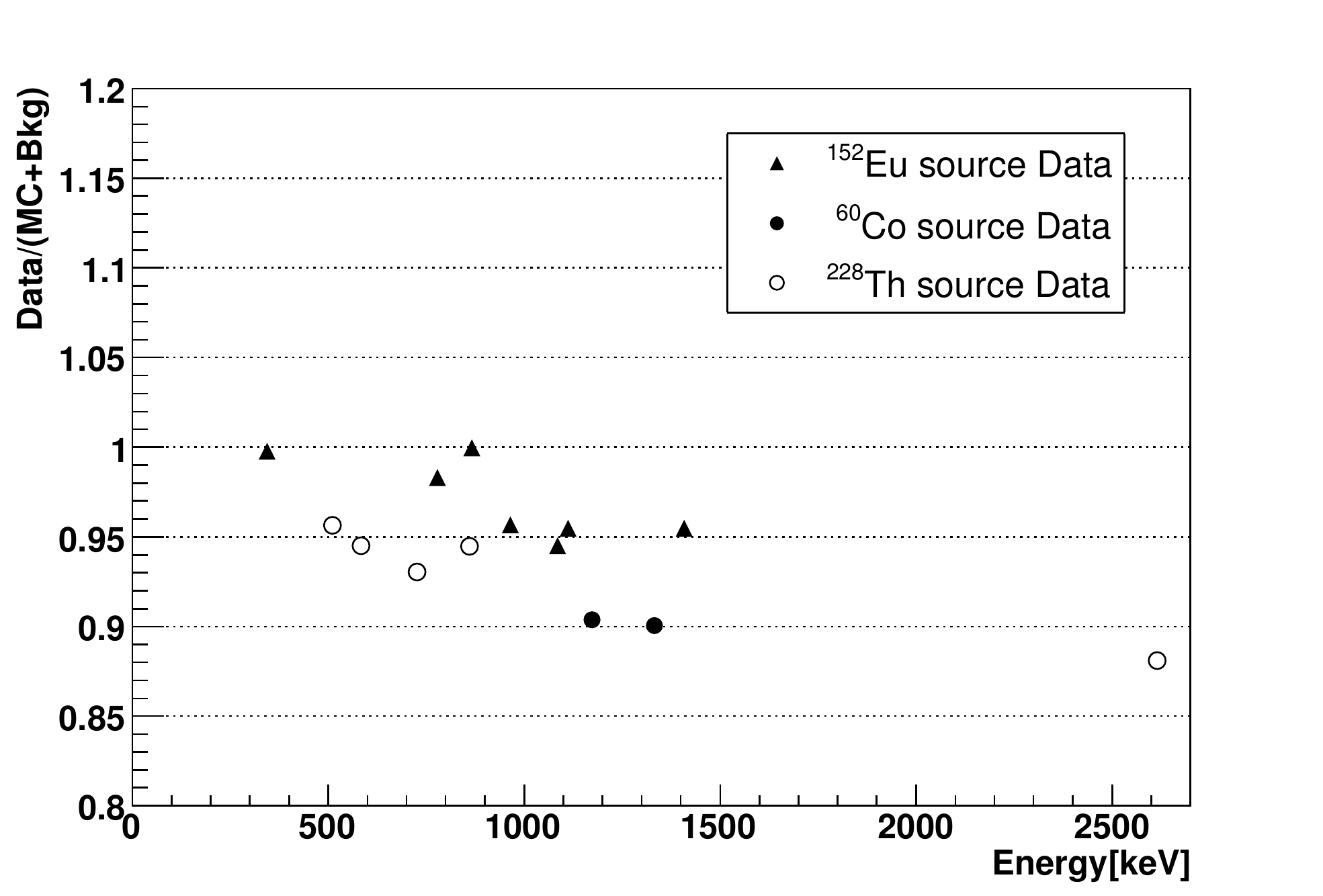} 
\caption{Ratio of number of events in the data and Monte Carlo plus
background samples for the most prominent peaks of $^{60}$Co (filled
circle), $^{152}$Eu (triangle) and $^{228}$Th (open circle). The statistical uncertainty is smaller than the marker size. 
\label{fig:DataOverMc}} 
\end{figure}

Fig.~\ref{fig:DataOverMc} shows the ratio of the number of events in
the data and Monte Carlo plus background samples for the
5$\sigma$-regions around the most prominent peaks. The markers
indicate the statistical uncertainty. The Monte Carlo plus background
data sets show a systematic excess of events which is slightly energy
dependent. The excess ranges from 1\% for an energy of 344.27~keV
($^{152}$Eu) up to 12\% at an energy of 2614.53~keV ($^{208}$Tl in the
$^{228}$Th decay chain). \\

The overall agreement between the experimental data and the prediction
from the simulation are on the 5-10\% level. This is reasonable agreement given
that the spectral shape and peak heights, which span four orders of magnitude in
intensity, depend sensitively on geometric effects. 
Discrepancies likely arise particularly from
\begin{itemize}
\item insufficient modeling of
the experimental setup. In Ref.~\cite{Hurtado:2004} it has been shown that an 
agreement better than a few percent can be obtained for $\gamma$-ray interactions 
in HPGe detectors with \textsc{Geant4} by a proper optimization of the 
model of the experimental setup; 
\item inefficiency effects which are not included in the
simulation, such as pile-up, trigger turn-on and digitization.
\end{itemize}

\section{Validation of the simulation}
\label{section:validation} 

The {\sc Geant4} simulation toolkit is used in various applications of
modern physics. These range from simulations in high-energy particle
physics to astrophysics and medical science. In parallel to the
development of new simulation modules the verification of the
simulation code is an important task for developers and users. Several
modules have been developed to describe the interactions of low energy
photons, electrons and hadrons with matter~\cite{geant4_lowe}. These
are of particular importance for applications such as {\sc MaGe} and
are tested within the two collaborations developing the software. In
the following, the current status of {\sc MaGe} verification efforts is
summarized. \\


The description of electromagnetic interactions in the energy region
up to several MeV was tested with high-purity germanium detector systems,
such as that presented in Section~\ref{section:example}. The most extensive
verification effort was performed with a 18-fold segmented \textsc{Gerda} prototype 
detector. The segmented germanium crystal was operated 
and exposed to several radioactive
sources ($^{60}$Co, $^{228}$Th, $^{152}$Eu)~\cite{Abt:2007rg,Abt:2007rf}. 
A large fraction of the emitted gammas deposit energy in
more than one segment. This feature allows 
these events to be distinguished from those which deposit energy in relatively
small volumes, such as $0\nu\beta\beta$-decays. 
Such segmentation-based discrimination between single- and 
multiple-site interactions was compared between
experiment and simulation, with deviations found on the
5\%~level. \\\

The description of neutron interactions with Ge is probed by comparing data
from a measurement of an AmBe source with predictions from {\sc
MaGe}. The measurements have been performed with a {\sc Clover}
detector and with the 18-fold segmented detector described
previously~\cite{siegfried_neutron}. At an energy level of several
MeV, neutrons mostly interact through elastic and inelastic scattering
as well as neutron absorption. The measured energy spectra were
studied and photon lines from neutron interactions with the germanium
detector itself and the surrounding materials were
identified~\cite{Mei:2007zd}. Several discrepancies in the {\sc
Geant4} simulation were identified. The 2223.0-keV peak from
H(n,$\gamma$)D appears at 2224.6 keV in {\sc Geant4} simulations (bug
report \# 955)~\cite{G4bug}.  This problem can be corrected by
modifying data files provided with {\sc Geant4}.  Meta-stable nuclear
states are not produced by {\sc Geant4} as a result of neutron
interactions (bug report \# 956)~\cite{G4bug}.  The {\sc Geant4}
collaboration is investigating this issue.  Neutrons also do not
produce internal conversion electrons in {\sc Geant4} (bug report \#
957)~\cite{G4bug}.  {\sc MaGe} developers are working towards a
solution for this problem.\\

The {\sc MaGe} package was also used to study and verify the
simulation of spallation neutron production and propagation. At the
CERN NA55 experiment the neutron production from a 190~GeV muon beam
incident on different targets was measured. At the SLAC electron beam
dump experiment the neutron propagation through different thicknesses of concrete was
measured. Both experiments were simulated within the {\sc MaGe}
framework.  It was found that \textsc{MaGe}/\textsc{Geant4} underestimate 
the neutron production from muon interactions measured by NA55, especially 
in high-Z materials, by more than a factor of two~\cite{mike_neutron}. 
Results obtained in the \textsc{MaGe} simulation of NA55 have been 
compared with the \textsc{Geant4}- and \textsc{Fluka}-based~\cite{fluka} 
Monte Carlo simulations of the same experiment performed in~\cite{Araujo:05}, and 
found to be consistent.
The disagreement between Monte Carlo simulation codes and NA55 data for muon-induced 
neutron production is discussed in detail in~\cite{Araujo:05,ilias06}. 
The attenuation of the neutron propagation is found to be larger in the simulation 
than measured in the SLAC experiment~\cite{mike_neutron}. A method to correct the neutron 
over-attenuation in \textsc{MaGe}-based simulations is described in~\cite{mike_neutron}. \\


\section{Conclusions}
\label{section:conclusions}

We presented the {\sc MaGe} framework for simulating interactions in
neutrinoless double-beta decay experiments that utilize enriched HPGe
detectors. The benefits of {\sc MaGe} can be summarized as:
\begin{itemize}
\item reliable Monte Carlo framework based on {\sc Geant4} for
low-background, low-energy experiments;
\item ongoing tests of the code and validation of the physics processes; 
\item flexible geometry and physics application that emphasizes code
reuse and verification; 
\item general purpose tools like surface 
and volume sampling, custom isotope decay generators, etc.
\end{itemize} 

In general there is good agreement between the {\sc MaGe} simulation
and the measurements of electromagnetic interactions with average
discrepancies of the order of (5-10)\%. Several problems have been
identified in the simulation of neutron interactions in {\sc Geant4}.
These problems have been reported to the {\sc Geant4} collaboration
and are under investigation by the {\sc MaGe} developers.
\\

We anticipate that {\sc MaGe} will form the foundation of the
simulation and analysis framework of the \textsc{Gerda} and \textsc{Majorana}
experiments. This framework is also applicable for other
low-background underground experiments, such as solar, 
reactor and geological neutrino experiments, direct dark matter searches, 
and other neutrinoless double-beta decay search.
These experiments share many detection techniques and background
issues in common with \textsc{Gerda} and \textsc{Majorana}.


\section*{Acknowledgments}

The authors would like to express their gratitude to all members of the 
\textsc{Majorana} and \textsc{Gerda} collaborations, for their continuous 
feedback and useful advices.
The authors would like to thank V.~Tretyak and C.~Tull for useful
discussions, and I.~Abt for useful
discussions and her efforts to unify the Monte Carlo of both
collaborations. This work has been supported by the
\textsc{ILIAS} integrating activity (Contract No.RII3-CT-2004-506222)
as part of the EU FP6 programme, by BMBF under grant 05CD5VT1/8, 
and by DFG under grant GRK 683.
This work was also supported by
Los Alamos National Laboratory's Laboratory-Directed Research and Development Program,
and by the Office of Science of the US Department of Energy 
at the University of Washington under Contract No. DE-FG02-97ER41020, 
at Lawrence Berkeley National Laboratory under Contract No. DE-AC02-05CH11231, 
and at Pacific Northwest National Laboratory under Contract No. DE-AC06-76RLO 1830. 
This research used the Parallel Distributed Systems Facility at the 
National Energy Research Scientific Computing Center, which is 
supported by the Office of Science of the U.S. Department of Energy 
under Contract No. DE-AC02-05CH11231.

\ifCLASSOPTIONcaptionsoff
  \newpage
\fi

\end{document}